\documentclass[prd,aps,showpacs,preprintnumbers,
nofootinbib,notitlepage,superscriptaddress]{revtex4-1}

\usepackage{amsmath}
\usepackage{amssymb}
\usepackage{graphicx}
\usepackage{epsfig}

\usepackage{epstopdf}
\usepackage{esint}

\begin{document}

\title{Bounds on majoron emission from muon to electron conversion experiments}
\author{Xavier \surname{Garcia i Tormo} }
\altaffiliation[Current address: ]{Institut f\"ur Theoretische Physik, Universit\"at Bern,
  Sidlerstrasse 5, CH-3012 Bern, Switzerland.}
\affiliation{Department of Physics, University of Alberta, Edmonton, Alberta,
Canada T6G 2G7}
\author{Douglas Bryman}
\affiliation{Department of Physics and Astronomy,
  University of British Columbia, Vancouver, British Columbia, Canada
  V6T 1Z1}
\author{Andrzej Czarnecki} 
\author{Matthew Dowling} 
\affiliation{Department of Physics, University of Alberta, Edmonton, Alberta,
Canada T6G 2G7}

\date{\today}

\preprint{Alberta Thy 09-11}

\begin{abstract}
In models where lepton number is considered to be a
spontaneously-broken global symmetry a massless Goldstone boson, the
majoron ($J$), appears. We describe a procedure to explore the
muon-electron-majoron coupling using the results from $\mu-e$
conversion search experiments. To accomplish that, we determine how the energy spectrum of the
muon decay into an electron and a majoron is modified by
binding effects in a muonic atom. We find that the
future $\mu \to e$ conversion experiments may be able to produce bounds on the
$\mu\to eJ$ rate which are comparable with the present ones from direct searches.
\end{abstract}


\maketitle

\section{Introduction}
The observation of neutrino oscillations has established that at
least some of the neutrinos have mass \cite{Schwetz:2011qt}. The origin of the small
neutrino mass differences and pattern of mixing angles which are seen
in the experiments is not known at present. Exploring the nature of
neutrino masses and mixing angles may allow us to glimpse particles
and interactions beyond the Standard Model \cite{GonzalezGarcia:2007ib}. One class of models that
generates neutrino masses, which could provide a possible
explanation of the origin of the observed oscillations, considers lepton number as a spontaneously-broken global symmetry. In that
case a massless Goldstone boson, the majoron ($J$) appears. Models of this kind were explored well before neutrino
oscillations were experimentally established
\cite{Chikashige:1980ui}. Some of those models
\cite{Gelmini:1980re,Aulakh:1982yn} predicted large additional
contributions to the invisible decay width of the $Z$ boson and were
excluded after the $Z$-width measurement at LEP. It is
possible, though, to construct supersymmetric models with spontaneous
breaking of $R$-parity (and thus of lepton number) where no significant invisible $Z$-width is
present, and the LEP bounds can be evaded
\cite{Masiero:1990uj}. 

In this paper, we focus on the model of Ref.~\cite{Masiero:1990uj} which permits
charged-lepton decays with majoron emission \cite{Romao:1991tp}. Those processes were
recently revisited in Ref.~\cite{Hirsch:2009ee}, where it was shown
that the $\mu\to e J$ decay rate is allowed to be large, and
could potentially be in a region where it could be measured. It is
quite interesting to study those decays, since they can explore
regions of the supersymmetric parameter space that are not probed by
collider searches. Experimentally it is quite a
difficult task to improve on the current limits for the branching
ratio $B(\mu\to e J)$. In
this paper we show that the future $\mu-e$ conversion experiments
\cite{LeptMomBook} may
be able to produce bounds that are comparable to the present ones. Previous studies of $(\mu^-,e^+)$ conversion mechanisms involved
majorons \cite{Zahir:1982km} in now disfavored models where the majoron is
a gauge non-singlet.

The current limit for the branching ratio of a muon decaying to an
electron and a majoron is \cite{BayesPhD}
\begin{equation}\label{eq:limBJ}
B({\mu}\to eJ)=\frac{{\Gamma}({\mu}\to eJ)}{{\Gamma}({\mu}\to e{\nu}_{\mu}\bar{{\nu}}_e)}<8.4\times 10^{-6}.
\end{equation}
We will study $\mu-e$ conversion experiments, which can reach high
sensitivities, and possibly improve on this limit. The $\mu-e$ conversion experiments produce
muonic atoms (by stopping muons in a target) and then search for the
process
\begin{equation}
\mu^{-}+(A,Z)\to e^{-}+(A,Z),\label{eq:mueconv}
\end{equation}
where $(A,Z)$ represents a nucleus of atomic number $Z$ and mass
number $A$.
The signal in conversion experiments is a
mono-energetic electron at energy $E_{{\mu}e}=m_{\mu}-E_{\textrm b}-E_{\textrm
  rec}$ (with $m_{\mu}$ the muon mass, $E_{\textrm b}$ the binding energy
of the muonic atom, and $E_{\textrm
  rec}$ the nuclear-recoil energy). If we now consider the
process of ${\mu}\to eJ$ decay in the orbit of a nucleus $(A,Z)$, i.e.
\begin{equation}\label{eq:emIO}
{\mu}^-+(A,Z)\to e^-+J+(A,Z),
\end{equation}
the outgoing electron can have energies up to the value of
the conversion
energy $E_{\mu e}$ since the nucleus can absorb momentum. $E_{\mu e}$ is also the maximum electron energy for muon
decay in orbit (DIO), which constitutes the main physics background source in
the search for the conversion process in Eq.~(\ref{eq:mueconv}). Since
the majoron $J$ is not observed, the signal for the process in
Eq.~(\ref{eq:emIO}) would be electrons below $E_{{\mu}e}$. Therefore,
the measurement of the electron spectrum close to $E_{{\mu}e}$, as is
done in the conversion experiments, can be used to obtain a limit for
$B({\mu}\to eJ)$. The electron spectrum for the decay in
Eq.~(\ref{eq:emIO}) is a delta-function like shape around $E_e\sim
m_{\mu}/2$ (where $E_e$ is the electron energy) with a tail due to
bound-state effects. The spectrum for the two-body free decay $\mu\to e J$ is just a
delta function at $E_e\sim m_{\mu}/2$. Even though we
only look for events at the high-energy tail of the
electron-energy spectrum in ${\mu}\to eJ$ decay in orbit, the fact that
the current limit in Eq.~(\ref{eq:limBJ}) is not very stringent,
combined with the high sensitivity that the future conversion experiments are
expected to reach, might allow them to improve on the present constraints.
The ${\mu}\to e$ conversion
experiments provide a limit on $B({\mu}\to eJ)$ by first determining
\begin{equation}
\frac{{\Gamma}({\mu}(A,Z)\to e(A,Z))}{{\Gamma}_{\textrm{capture}}}=:R_{{\mu}e},
\end{equation}
where ${\Gamma}_{\textrm{capture}}$ is the rate of the nuclear muon
capture, ${\Gamma}_{\textrm{capture}}=\Gamma\left(\mu+(A,Z)\to(A,Z-1)+\nu_{\mu}\right)$. This
result places a bound on the majoron emission in muon decay
according to
\begin{equation}
\frac{{\Gamma}({\mu}\to eJ)\times f_J}{{\Gamma}_{\textrm{capture}}}\sim N_RR_{{\mu}e},
\end{equation}
where $f_J$ is the fraction of ${\mu}\to eJ$ decay in orbit events in the
signal region of the conversion experiment and  $N_R$ is a correction
factor for the phase space region used in the search for
$\mu+(A,Z)\to e+J+(A,Z)$. The limit on the
branching ratio is then given by
\begin{equation}\label{eq:limBR}
B({\mu}\to eJ)=\frac{{\Gamma}({\mu}\to eJ)}{{\Gamma}({\mu}\to
  e{\nu}_{\mu}\bar{{\nu}}_e)}\sim\frac{N_RR_{{\mu}e}}{f_J}\frac{{\Gamma}_{\textrm{capture}}}{{\Gamma}({\mu}\to e{\nu}_{\mu}\bar{{\nu}}_e)}.
\end{equation}
In the near future, the DeeMe
Collaboration \cite{DeeMe} has proposed to reach $R_{{\mu}e}\sim
10^{-14}$ sensitivity. Furthermore, the planned conversion experiments, Mu2e at Fermilab
\cite{Carey:2008zz} and COMET at J-PARC \cite{Cui:2009zz}, aim for sensitivities at the $R_{{\mu}e}\sim 10^{-16}$ level; in addition,  both
experiments are speculating on a later phase which would aim for a
sensitivity of $10^{-18}$. Comparing Eq.~(\ref{eq:limBR}) with the
present limit in Eq.~(\ref{eq:limBJ}) we can examine the potential
of those experiments to improve on the present bounds, as long as the
factor $(f_J{\Gamma}({\mu}\to
e{\nu}_{\mu}\bar{{\nu}}_e))/ (\Gamma_{\textrm{capture}}N_R)$ is not much smaller than $10^{-11}$.

The paper is organized as follows: in Sec.~\ref{sec:MEIO} we compute the electron
spectrum for the $\mu\to e J$ decay in atomic orbit. Then in
Sec.~\ref{sec:bounds}, we present our numerical results and discuss in detail the
bounds that we can obtain for the branching ratio $B(\mu\to eJ)$. We
conclude in Sec.~\ref{sec:concl}.

\section{Majoron emission in orbit}\label{sec:MEIO}
Our aim is to calculate the electron spectrum for majoron emission ${\mu}\to eJ$ in orbit
(MEIO), i.e. the process in Eq.~(\ref{eq:emIO}).

We write the interaction that mediates majoron emission in muon decay ${\mu}\to eJ$ as
\begin{equation}
\mathcal{L}=\bar{{\mu}}g_1P_ReJ+\bar{{\mu}}g_2P_LeJ,
\end{equation}
where $P_R=(1+{\gamma}_5)/2$ and $P_L=(1-{\gamma}_5)/2$. The couplings $g_1$ and
$g_2$ are dimensionless, i.e. they can be written in terms of parameters of the
underlying supersymmetric model \cite{Hirsch:2009ee}. The matrix element for the free $\mu\to
e J$ decay is given by
\begin{equation}
\mathcal{M}=\bar{u}(p_e)\left(g_1P_R+g_2P_L\right)u(p_{\mu})\quad;\quad|\mathcal{M}|^2=\bar{u}(p_e)\left(g_1P_R+g_2P_L\right)u(p_{\mu}) \bar{u}(p_{\mu})\left(g_1P_L+g_2P_R\right)u(p_e),
\end{equation}
where $p_{\mu}$ and $p_e$ are the momentum of the muon and the electron,
respectively. Summing over the electron spin and averaging over the muon spin
we obtain
\begin{equation}
\overline{\sum}|\mathcal{M}|^2=(g_1^2+g_2^2)p_e\cdot p_{\mu},
\end{equation}
where we considered the electron to be massless. We always consider a
massless electron in this paper, since we are interested in the high-energy part of the spectrum for MEIO. The free decay rate, ${\Gamma}_0$, is then given by
\[
{\Gamma}_0=\frac{1}{2m_{\mu}}\int\frac{d^3p_e}{(2{\pi})^32E_e}\frac{d^3p_J}{(2{\pi})^32E_J}(2{\pi})^4{\delta}^{(4)}(p_{\mu}-p_e-p_J)
\overline{\sum}|\mathcal{M}|^2
\]
\begin{equation}\label{eq:freeJdec}
=\frac{g_1^2+g_2^2}{16{\pi}}\int dE_e\frac{E_e^2}{m_{\mu}-E_e}{\delta}\left(E_e-\frac{m_{\mu}}{2}\right)=\frac{(g_1^2+g_2^2)m_{\mu}}{32{\pi}},
\end{equation}
where $p_J=(E_J,\vec{p}_J)$ is the majoron 4-momentum. Obviously, since this is a 2-body decay, the electron-energy spectrum is just
a $\delta$ function at $E_e=m_{\mu}/2$. For MEIO
Eq.~(\ref{eq:freeJdec}) gets replaced by
\begin{equation}\label{eq:gammbd}
{\Gamma}=\sum_{e^-\textrm{spin}}\int\frac{d^3p_e}{(2{\pi})^32E_e^2}\frac{d^3p_J}{(2{\pi})^32E_J}(2{\pi}){\delta}(E_{\mu}-E_e-E_J)\mathcal{J}\mathcal{J}^{\dagger},
\end{equation}
where $E_{\mu}=m_{\mu}-E_{\textrm b}$, and
\begin{equation}\label{eq:curr}
\mathcal{J}:=\int d^3re^{-i\vec{p}_J\cdot\vec{r}}\bar{\varphi}_e(g_1P_R+g_2P_L){\varphi}_{\mu},
\end{equation}
where $\varphi_{e}$
and $\varphi_{\mu}$ are the solutions of the Dirac equation (in the
potential created by the nucleus) for the electron and the muon,
respectively. We incorporate the average over the muon spin in the
definition of $\varphi_{\mu}$,
while we do not incorporate the sum over the electron spin in the
definition of $\varphi_{e}$.  When the muonic atom is formed the muon cascades down almost immediately
to the ground state, and this process also depolarizes the muon \cite{:2009yn}.
We consider an unpolarized muon in the
$1S$ state and write the muon wavefunction as
\begin{equation}
\varphi_{\mu}(\vec{r})=\sum_sa_s \left(\begin{array}{c}G\chi_{-1}^s\\iF\chi_{1}^s\end{array}\right),
\end{equation}
where $a_s$ is the amplitude of the muon state with spin
projection $s$. For an unpolarized muon we have
$|a_s|^2=1/2$. $\chi_{\kappa}^{\mu}=\chi_{\kappa}^{\mu}(\hat{r})$ are
the spin-angular functions, which are given by
\begin{equation}\label{eq:chikm}
\chi_{\kappa}^{\mu}=\sum_mC\left( l\frac{1}{2}j;\mu-m\, m\, \mu\right)Y_{l}^{\mu-m}\chi^m,
\end{equation}
with $C(lsj;l_zs_zj_z)$ the Clebsch-Gordan
coefficients, $Y_l^{\mu}$ are the spherical harmonics and $\chi^m$ are the spin
$1/2$ eigenfunctions. $G$ and $F$ are the solutions of the radial
Dirac equations, which are taken to be normalized as 
\begin{equation}\label{eq:munorm}
\int r^2(F^2+G^2)dr=1.
\end{equation}
The electron wavefunction is expanded in partial waves according to
\begin{equation}
\varphi_e(\vec{r})=\sum_{\kappa\mu}a_{\kappa\mu t}\psi_{\kappa}^{\mu}=\sum_{\kappa\mu}a_{\kappa\mu t}\left(\begin{array}{c}g_{\kappa}\chi_{\kappa}^{\mu}\\if_{\kappa}\chi_{-\kappa}^{\mu}\end{array}\right),
\end{equation}
where $g_{\kappa}$ and $f_{\kappa}$ are the solutions of the radial
Dirac equations for the electron, labeled by $\kappa = \pm 1, \pm 2, \ldots$ \cite{Rosebook}, $t$ is the $z$-component of the electron spin, and the $a_{\kappa\mu t}$
coefficients are given by
\begin{equation}\label{eq:apwe}
a_{\kappa\mu t}=i^{l_{\kappa}}\frac{4\pi}{\sqrt{2}}C\left(l_{\kappa}\frac{1}{2}j_{\kappa};\mu-t\, t\,
  \mu\right)Y_{l_{\kappa}}^{\mu-t\, *}(\hat{p}_e)e^{-i\delta_{\kappa}},
\end{equation}
where $\delta_{\kappa}$ is the Coulomb phase shift (the distortion from a
plane wave due to the potential of the nucleus),
$j_{\kappa}=|\kappa|-1/2$, and $l_{\kappa}=j_{\kappa}-\textrm{Sign}(\kappa)/2$. The electron
wavefunctions are normalized in the energy scale, according to
\begin{equation}\label{eq:enorm}
\int d^3r\psi_{\kappa,W}^{\mu*}\psi_{\kappa',W'}^{\mu'}=2\pi\delta_{\mu\mu'}\delta_{\kappa\kappa'}\delta(W-W'),
\end{equation}
where $\psi_{\kappa,W}^{\mu}$ corresponds to a solution with energy
$W$. 

We ignored nuclear-recoil effects to write Eq.~(\ref{eq:gammbd}) but will incorporate them later. Integrating over the solid angle in Eq.~(\ref{eq:curr}) we obtain
\begin{eqnarray}
\mathcal{J} & = & \frac{\sqrt{4{\pi}}}{2}\sum_{KM}(-i)^K\sum_{s{\kappa}{\mu}}a^*_{{\kappa}{\mu}t}a_s\int
dr
r^2j_K(p_Jr)C\left(K\frac{1}{2}j_{\kappa};Ms{\mu}\right)Y_K^{M*}(\hat{p}_J)\nonumber\\
&&\times\left\{(g_1+g_2)\left(g_{\kappa}G-f_{\kappa}F\right){\delta}_{Kl_{\kappa}}+i (g_1-g_2)\left(f_{\kappa}G+g_{\kappa}F\right){\delta}_{Kl_{-{\kappa}}}\right\}.
\end{eqnarray}
We can then write the result for the spectrum,
\begin{equation}\label{eq:spec}
\frac{1}{{\Gamma}_0}\frac{d{\Gamma}}{dE_e}=\sum_{K{\kappa}}\frac{1}{{\pi}m_{\mu}}(2j_{\kappa}+1)[E_{\mu}-E_e]\left|\mathcal{S}_{K{\kappa}}\right|^2,
\end{equation}
where
\begin{equation}\label{eq:S}
\mathcal{S}_{K{\kappa}}:=\int dr r^2 j_K\left([E_{\mu}-E_e]r\right)\left\{\left(g_{\kappa}G-f_{\kappa}F\right){\delta}_{Kl_{\kappa}}+i \left(f_{\kappa}G+g_{\kappa}F\right){\delta}_{Kl_{-{\kappa}}}\right\}.
\end{equation}
In Eq.~(\ref{eq:spec}), the sum over $K$ goes from 0 to $\infty$, and for a
given value of $K$, $\kappa$ can only take the values $\pm K$ and
$\pm(K+1)$, but $\kappa$ can never be equal to 0. $j_n(z)$ is the
spherical Bessel function of order $n$.

Since future $\mu \to e$ conversion experiments may use aluminum or
heavier nuclei as targets, the nucleus is, at least, more than 200 times heavier than
the muon and nuclear-recoil effects are negligible for most of the
electron spectrum. Recoil effects could be important close to the
high-energy endpoint since they modify the maximum allowed electron
energy. In that region we can approximate the nuclear-recoil energy
$E_{\textrm rec}$ as \cite{Shanker:1981mi}
\begin{equation}
E_{\textrm rec}=\frac{|\vec{p}_N|^2}{2m_N}\simeq \frac{E_e^2}{2m_N},
\end{equation}
with $m_N$ the mass of the nucleus and $\vec{p}_N$ its
three-momentum. Within this approximation, the inclusion of recoil
effects reduces to the substitution 
\begin{equation}
E_{\mu}-E_e\to E_{\mu}-E_e-\frac{E_e^2}{2m_N},
\end{equation}
inside the square brackets in Eqs.~(\ref{eq:spec}) and
(\ref{eq:S}). The endpoint is now at
$E_e=E_{\mu}-E_{\mu}^2/(2m_N)$. Inclusion of nuclear-recoil
effects beyond the approximation in the equations above is
unnecessary for all practical purposes. Our final result for the electron spectrum in MEIO then
reads
\begin{equation}\label{eq:specrec}
\frac{1}{{\Gamma}_0}\frac{d{\Gamma}}{dE_e}=\sum_{K{\kappa}}\frac{1}{{\pi}m_{\mu}}(2j_{\kappa}+1)\left[E_{\mu}-E_e-\frac{E_e^2}{2m_N}\right]\left|\mathcal{S}_{K{\kappa}}\right|^2,
\end{equation}
with
\begin{equation}\label{eq:Srec}
\mathcal{S}_{K{\kappa}}:=\int dr r^2 j_K\left(\left[E_{\mu}-E_e-\frac{E_e^2}{2m_N}\right]r\right)\left\{\left(g_{\kappa}G-f_{\kappa}F\right){\delta}_{Kl_{\kappa}}+i \left(f_{\kappa}G+g_{\kappa}F\right){\delta}_{Kl_{-{\kappa}}}\right\}.
\end{equation}

\subsection{Taylor expansion around the endpoint}\label{subsec:TE}
We can perform a Taylor expansion of the electron energy spectrum in Eq.~(\ref{eq:specrec})
around the endpoint to make its behavior manifest. For that, we need the Taylor expansion of
$\mathcal{S}_{K{\kappa}}$ in Eq.~(\ref{eq:Srec}). We obtain, noting that the electron
wavefunctions $g_{\kappa}$ and $f_{\kappa}$ depend on $E_e$,
\begin{eqnarray}\label{eq:TeSrec}
\mathcal{S}_{K{\kappa}} & = & \int dr
r^2\left(g_{-1}G-f_{-1}F\right){\delta}_{0K}{\delta}_{-1{\kappa}}\nonumber\\
&&+\int dr
r^2\left\{\frac{r}{3}\left(1+\frac{E_{\mu}}{m_N}\right)\left[i\left(f_{-1}G+g_{-1}F\right) {\delta}_{1K}{\delta}_{-1{\kappa}}+\left(g_{-2}G-f_{-2}F\right) {\delta}_{1K}{\delta}_{-2{\kappa}}\right]\right.\nonumber\\
&&-\left(g_{-1}'G-f_{-1}'F\right) {\delta}_{0K}{\delta}_{-1{\kappa}}\Big\}\left(E_{\mu}-E_e-\frac{E_e^2}{2m_N}\right)+\mathcal{O}\left(\left(E_{\mu}-E_e-\frac{E_e^2}{2m_N}\right)^2\right),
\end{eqnarray}
where 
\begin{equation}
f_{\kappa}':=\frac{df_{\kappa}}{dE_e}\quad ;\quad g_{\kappa}':=\frac{dg_{\kappa}}{dE_e}.
\end{equation}
It is understood that the electron wavefunctions in
Eq.~(\ref{eq:TeSrec}) (or their
derivatives) are evaluated at the endpoint energy
$E_e=E_{\mu}-E_{\mu}^2/(2m_N)$. The corresponding Taylor expansion
without including recoil effects is recovered by taking the limit
$m_N\to\infty$. Since we have the property
\begin{equation}
\int dr r^2 g_{-1}G=\int dr r^2 f_{-1}F,
\end{equation}
when the energies of the muon and the electron are equal \cite{Shanker:1981mi}, the leading
term in the expansion of $\mathcal{S}_{K\kappa}$ vanishes when we
neglect nuclear-recoil effects. Therefore, the Taylor expansion of the
electron spectrum without nuclear-recoil effects starts at order
$(E_{\mu}-E_e)^3$. If we take nuclear recoil into account the leading
term of the Taylor expansion for $\mathcal{S}_{K\kappa}$ no longer vanishes exactly, but it is suppressed by
the inverse of $m_N$.

\section{Bounds on the branching ratio}\label{sec:bounds}
Equation (\ref{eq:specrec}) gives the electron
spectrum for MEIO. We will use it here to obtain bounds for
the branching ratio $B(\mu\to eJ)$. We solve the Dirac equation
numerically to obtain the electron ($g_{\kappa}$ and $f_{\kappa}$) and
muon ($G$ and $F$) wavefunctions that appear in
Eq.~(\ref{eq:specrec}). To do that, we consider a nucleus of finite size, characterized by a two-parameter
Fermi distribution $\rho(r)$, given by 
\begin{equation}
\rho(r)=\rho_{0}\frac{1}{1+e^{\frac{r-r_{0}}{a}}}.\label{eq:2pFd}
\end{equation}
The parameters $r_0$ and $a$, the nuclear masses, the muon binding energy
$E_\mu$, and the endpoint energy $E_{\mu e}$ for the elements of
current experimental interest are  summarized in Table \ref{tab:pars}.
The normalization factor  $\rho_0$ in Eq. (29) is determined from the condition $\int d^3r\, \rho(r) = -Z\alpha$. 
\begin{table}
\caption{Values for the parameters in the Fermi distribution in
  Eq.~(\ref{eq:2pFd}) \cite{De Jager:1987qc}, nuclear masses, muon
  energy $E_{\mu}$, and endpoint energy $E_{\mu e}$,
  for the elements used in the text.}\label{tab:pars}
\begin{ruledtabular}
\begin{tabular}{cccccc}
Nucleus & $r_0$ (fm) & $a$ (fm) & $m_N$ (MeV) & $E_{\mu}$ (MeV) &$E_{\mu e}$ (MeV) \\
\hline
Al($Z=13$) & 2.84 & 0.569 & 25133 & 105.194 &104.973 \\
Ti($Z=22$) & 3.84 & 0.588 & 44588 & 104.394 &104.272 \\
Au($Z=79$) & 6.38 & 0.535 & 183473 & 95.533 & 95.508
\end{tabular}
\end{ruledtabular}
\end{table}
For the muon mass and the fine
structure constant we use the values $m_{\mu}=105.6584\,\textrm{MeV},
\alpha=\frac{1}{137.036}$. Radiative corrections were not included in
Eq.~(\ref{eq:specrec}) but are not expected to significantly
modify the results. Note that, as it happens for the usual
muon DIO (see Ref.~\cite{muonDIO}), to obtain the correct result for
the high-energy tail of the spectrum it is necessary to include finite
nuclear size effects, the interaction of the outgoing electron with
the field of the nucleus, the Dirac (rather than Schr\"odinger)
wavefunction for the muon and (at least for not very heavy elements)
nuclear-recoil effects, as we did.

Conversion experiments measure the electron spectrum in a window of a
few MeV near the energy $E_{\mu e}$, in search for a peak that would
reveal the conversion process in Eq.~(\ref{eq:mueconv}). In this
energy window, electrons from muon DIO are also present and are seen
by the experiments. The spectrum for muon DIO has been recently
studied in detail \cite{muonDIO} and it is now under good
theoretical control. Electrons coming from MEIO would appear as an
additional contribution on top of the electrons coming from muon DIO. 

\subsection{Existing conversion results}
\begin{figure}
\centering
\includegraphics[width=8cm]{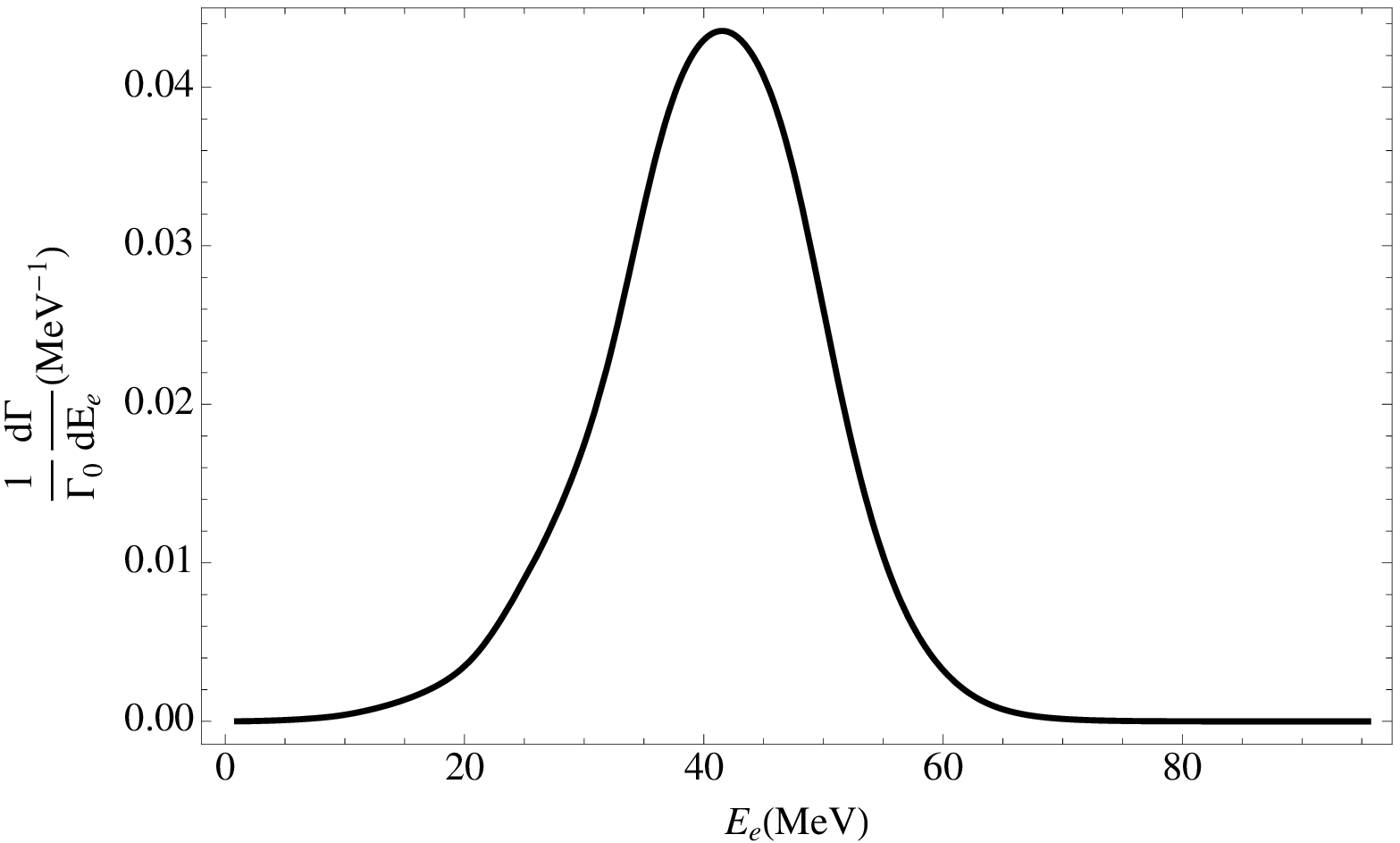}\hspace{.4cm}\includegraphics[width=8cm]{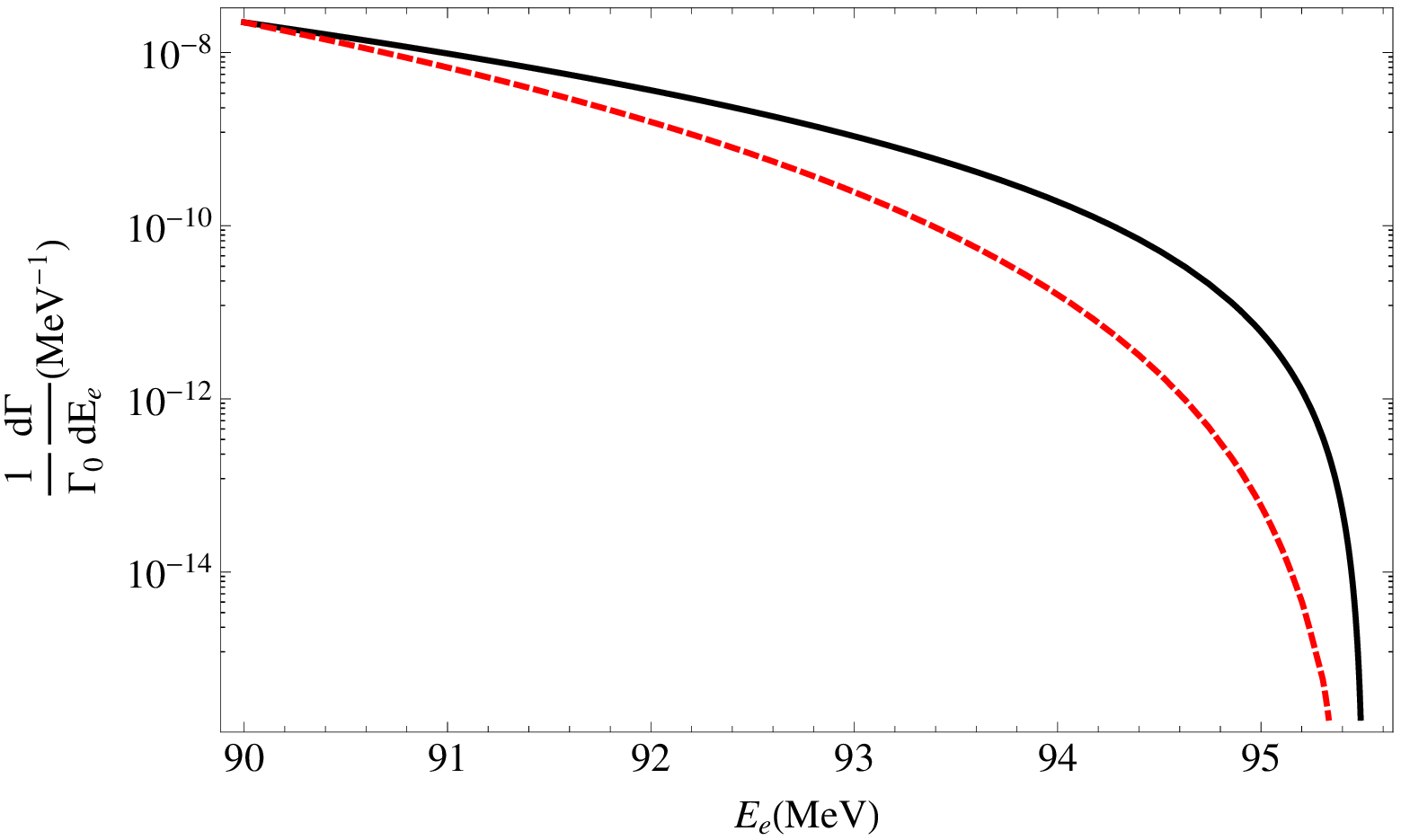}
\caption{Electron spectrum for majoron emission in orbit for
  gold (solid line). The second panel is a zoom for $E_e>90$
  MeV. The dashed line in the second panel is the electron spectrum for DIO in gold, multiplied by a constant ($C=333$) to make it coincide with the
MEIO rate at $E_e=90$ MeV.}\label{fig:specAu}
\end{figure}

Currently the most stringent upper limit for the conversion branching
ratio is given by the SINDRUM II Collaboration, $R_{\mu e}<7\times
10^{-13}$ ($90\%$ C.L.), using a gold (Au, $Z=79$) target \cite{Bertl:2006up}. Ref.~\cite{Bertl:2006up} measured the electron spectrum in a region
from 90 MeV to $E_{\mu e}$. From
Eq.~(\ref{eq:specrec}) we obtain the electron spectrum for MEIO for
gold, which for illustration we plot in Fig.~\ref{fig:specAu} as the
solid line. 
We find that the MEIO spectrum
for energies $E_e>90$ MeV is very well fitted by 
\begin{equation}\label{eq:fitepAu}
\frac{1}{{\Gamma}_0}\left.\frac{d{\Gamma}}{dE_e}\right|_{{\textrm
    Au,}\, E_e>90\textrm{ MeV}}=\frac{1}{m_{\mu}}\left( 5.292\times 10^{-3}\delta^3+
9.629\times 10^{-2}{\delta}^4+1.125 \,{\delta}^5+22.94 \,{\delta}^6\right),
\end{equation}
where
\begin{equation}
\delta:=\frac{E_{\mu}-E_e-\frac{E_e^2}{2m_N}}{m_{\mu}}.
\end{equation}
Since the gold nucleus is quite heavy, the terms proportional to $\delta$
and $\delta^2$ in the spectrum are found to be negligibly small (see
the discussion in Sec.~\ref{subsec:TE}). The fraction $f_J$ of $\mu\to eJ$ decay in orbit events for energies above
90 MeV is given by
\begin{equation}
\left.f_J\right|_{{\textrm
    Au,}\, E_e>90\textrm{ MeV}}=\int_{90\textrm{MeV}}^{E_{\mu e}}\frac{1}{{\Gamma}_0}\frac{d{\Gamma}}{dE_e}dE_e=2.4\times 10^{-8}.
\end{equation}
The total muon lifetime in gold is 88 ns, determined by the capture rate. Therefore,
using the estimate for the limit on the branching ratio that we
derived in Eq.~(\ref{eq:limBR}), we have
\begin{equation}\label{eq:limAuasR}
B({\mu}\to eJ)\sim\frac{N_RR_{{\mu}e}}{f_J}\frac{{\Gamma}_{\textrm{capture}}}{{\Gamma}({\mu}\to
  e{\nu}_{\mu}\bar{{\nu}}_e)}\sim\frac{N_RR_{{\mu}e}}{f_J}\frac{2.197\, {\mu}s}{88\,
  ns}\sim\frac{N_RR_{{\mu}e}}{f_J}\, 25,
\end{equation}
where we also used that the free muon width is ${\Gamma}({\mu}\to
e{\nu}_{\mu}\bar{{\nu}}_e)=1/(2.197{\mu}s)$, and $N_R$ is the factor representing the event distribution observed below the
conversion peak. The upper limit $R_{{\mu}e}<7\times 10^{-13}$
set by Ref.~\cite{Bertl:2006up} was obtained from a likelihood analysis considering the mono-energetic
conversion signal, muon DIO, and additional backgrounds from radiative
muon capture, pion decays and cosmic rays. The shape of the electron energy distribution
seen in the experiment was well described by muon DIO. It is
reasonable to assume that the events seen in the region $E_e>90$ MeV came from muon DIO and that no
additional events from majoron emission were present. Nevertheless, to
avoid doing background subtractions of DIO events, we will consider that
the number of MEIO events cannot be larger than those seen in the
experiment. Six events
were seen for $E_e>90$ MeV, and if we assume a Gaussian
distribution (with mean and variance equal to 6), the $90\%$
C.L. limit corresponds to 9 events. Therefore, to
obtain the limit on $B({\mu}\to eJ)$, we use Eq.~(\ref{eq:limAuasR}) with the factor $N_R=9/2.3$ (where
the 2.3 in the denominator corresponds to the events for $90\%$ C.L. on a
Poisson distribution with expected number of occurrences 1 or 0). Doing
that we obtain
\begin{equation}\label{eq:limfrSII}
B({\mu}\to eJ)\lesssim\frac{N_RR_{{\mu}e}}{f_J}\, 25\sim\frac{\frac{9}{2.3}7\times 10^{-13}}{2.4\times 10^{-8}}\, 25\sim 3\times 10^{-3}.
\end{equation}
The upper limit in Eq.~(\ref{eq:limfrSII}) is still less restrictive than
the current limit in Eq.~(\ref{eq:limBJ}). Thus, the existing conversion
data does not improve on the present limits for the $\mu-e-J$ coupling. 

\subsection{New conversion experiments (Mu2e and COMET)}
Future $\mu \to e$ conversion experiments Mu2e at Fermilab \cite{Carey:2008zz} and COMET at J-PARC
\cite{Cui:2009zz} aim to reach sensitivities at the
$10^{-16}$ level, and to use an aluminum (Al, $Z=13$) target. The
ratio of muon capture width in Al over the free muon width is around $1.5$ (as
compared to 25 for Au, as in Eq.~(\ref{eq:limAuasR})), but $f_J$ is lower due to the much
lower $Z$. The $\mu \to e$ conversion energy in Al is around 105 MeV (see
Table \ref{tab:pars}). Using
$E_e>100$ MeV as the signal region, $f_J$ for Al is $2.2\times
10^{-10}$, and $N_R=27$ based on subtraction of the DIO background, resulting in
\begin{equation}
B({\mu}\to eJ)\sim\frac{N_RR_{{\mu}e}}{f_J}\frac{{\Gamma}_{\textrm{capture}}}{{\Gamma}({\mu}\to
  e{\nu}_{\mu}\bar{{\nu}}_e)}\sim\frac{N_RR_{{\mu}e}}{f_J}1.5\sim\frac{27\times
  10^{-16}}{2.2\times
  10^{-10}}1.5\sim 1.9\times 10^{-5},
\end{equation}
which is comparable to the current limit in
Eq.~(\ref{eq:limBJ}).  If a sensitivity of $10^{-18}$ is reached, the limit on $B({\mu}\to eJ)$ will improve by an order of magnitude due to improved statistical precision on DIO.

\begin{figure}
\centering
\includegraphics[width=8cm]{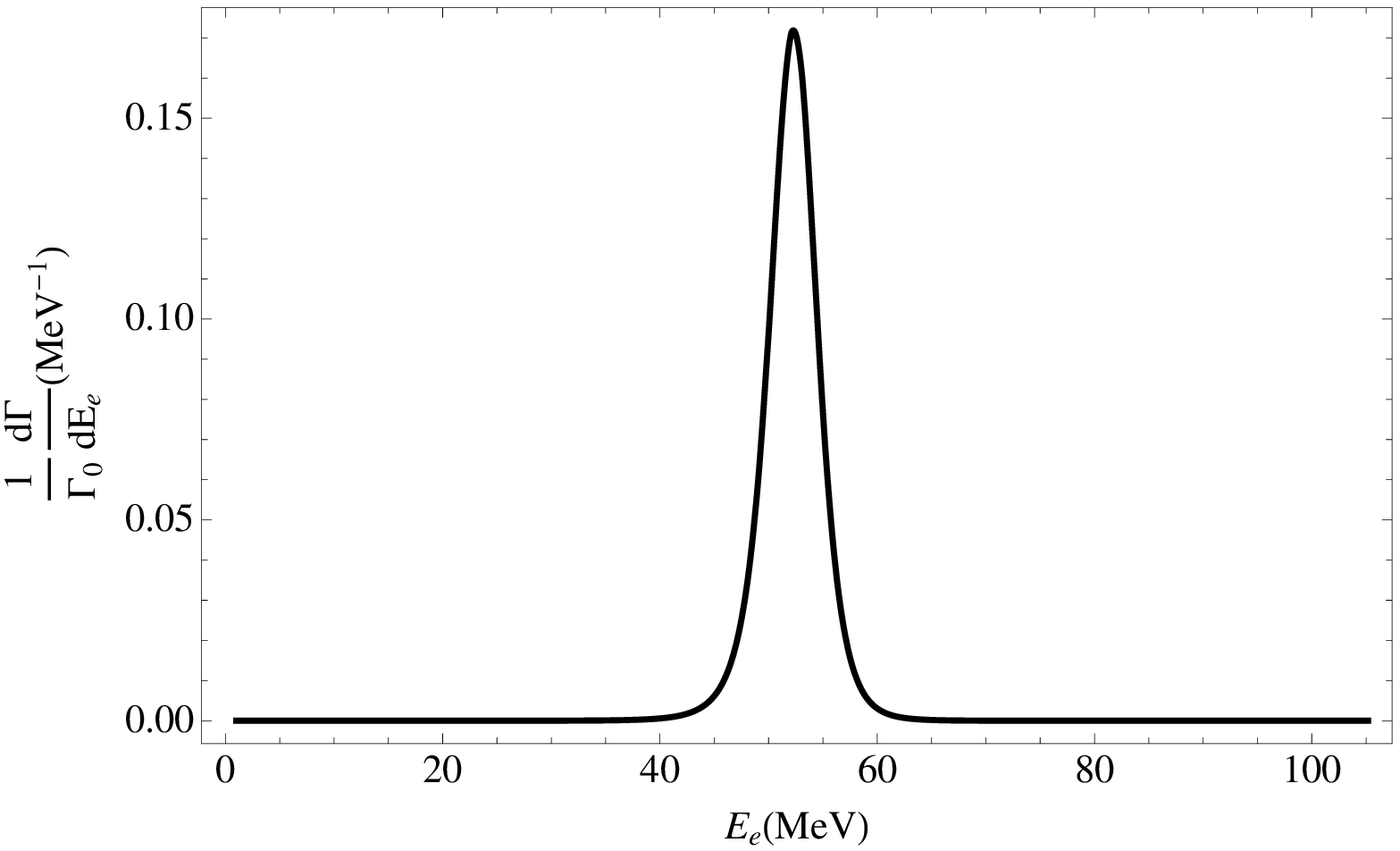}\hspace{.4cm}\includegraphics[width=8cm]{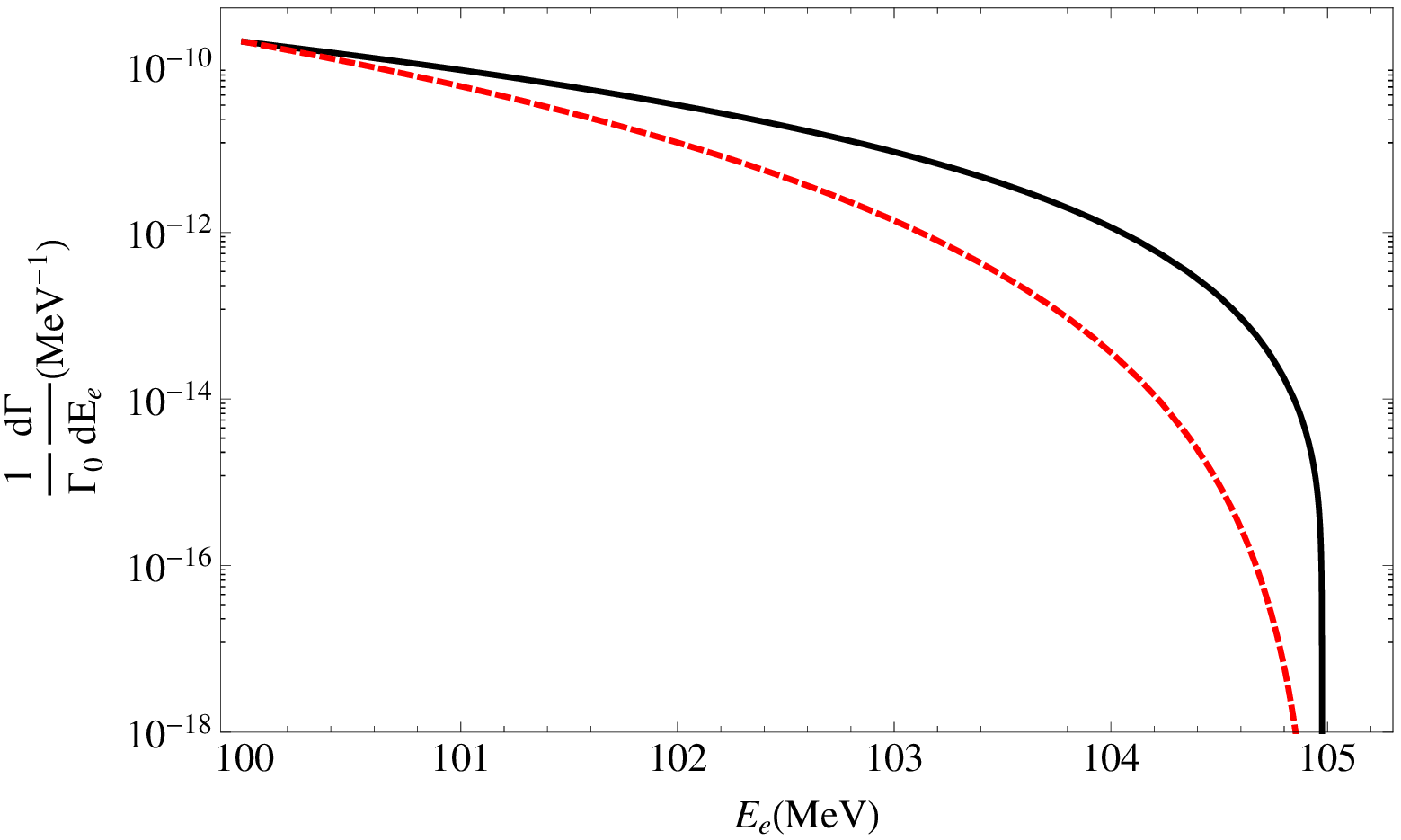}
\caption{Electron spectrum for majoron emission in orbit for
  Al (solid line). The second panel is a zoom for
  $E_e>100$ MeV. The dashed line in the second panel is the electron
  spectrum for DIO in Al, multiplied by a constant ($C=415$) to make it coincide with the
MEIO rate at $E_e=100$ MeV.}\label{fig:specAl}
\end{figure}

To obtain a more detailed estimate of the upper limit for
$B({\mu}\to eJ)$ that Mu2e and COMET could obtain, we need to convolute
the MEIO spectrum with the experimental energy resolution modeled by those collaborations.
Once again, from Eq.~(\ref{eq:specrec}) we get the electron spectrum for MEIO for Al, which for
illustration we plot in Fig.~\ref{fig:specAl} as the solid line.
We find that the spectrum
for energies $E_e>100$ MeV is very well fitted
by\footnote{Eq.~(\ref{eq:fitepAl}) may  be used in a wider
  energy range. It reproduces Eq.~(\ref{eq:specrec}) for
  Al with an accuracy better than $10\%$ until $E_e\sim 90$ MeV.}
\[
\frac{1}{{\Gamma}_0}\left.\frac{d{\Gamma}}{dE_e}\right|_{{\textrm
    Al,}\, E_e>100\textrm{ MeV}}
\]
\begin{equation}\label{eq:fitepAl}
=\frac{1}{m_{\mu}}\left( 3.289\times
  10^{-10}\delta +3.137\times 10^{-7}\delta^2+ 1.027\times 10^{-4}\delta^3+
1.438\times 10^{-3}{\delta}^4+2.419 \times 10^{-3}{\delta}^5+1.215\times 10^{-1} \,{\delta}^6\right).
\end{equation}
Convoluting the above equation with the estimated energy resolution of the
experiment (which we denote as $F(E_e)$) we can obtain a value of $f_J$
\begin{equation}
\left.f_J\right|_{E_e>x}=\int_{x}^{E_{\mu
    e}}\left(\frac{1}{{\Gamma}_0}\frac{d{\Gamma}}{dE_e}\otimes F\right)dE_e,
\end{equation}
where $\otimes$ denotes the convolution.  
Using the estimated energy resolution for Mu2e and COMET, and
estimating the expected number of DIO events in the signal region
($N_{\mathrm{DIO}}\simeq2400$), we
find that the bound on $B({\mu}\to eJ)$ that a conversion experiment
with $R_{\mu e}\sim 10^{-16}$ on Al may be able to place will be
$B({\mu}\to eJ)<2\times 10^{-5}$, roughly at the same level as the current one in Eq.~(\ref{eq:limBJ}).

Since COMET and Mu2e also consider Ti as a viable target we give, for
completeness, the polynomial that fits the MEIO spectrum in Ti for
$E_e>99$ MeV \footnote{It can be used until $E_e\sim 93$ MeV to reproduce
Eq.~(\ref{eq:specrec}) for Ti with an accuracy better than $10\%$.}
\[
\frac{1}{{\Gamma}_0}\left.\frac{d{\Gamma}}{dE_e}\right|_{{\textrm
    Ti,}\, E_e>99\textrm{ MeV}}
\]
\begin{equation}\label{eq:fitepTi}
=\frac{1}{m_{\mu}}\left( 5.404\times
  10^{-10}\delta +9.301\times 10^{-7}\delta^2+ 5.552\times 10^{-4}\delta^3+
8.113\times 10^{-3}{\delta}^4+5.470 \times 10^{-2}{\delta}^5+4.244\times 10^{-1} \,{\delta}^6\right).
\end{equation}
The bounds on $B({\mu}\to eJ)$ we would obtain with a Ti target
are similar to those for Al.

\subsection{Discussion}
We have shown in the previous section that a $\mu \to e$ conversion experiment with an Al target
and sensitivity at the $10^{-16}-10^{-18}$ level may be able to
produce bounds on the $\mu-e-J$ coupling which are competitive with
the present ones. If the electron
distribution seen in the experiments around $E_{\mu e}$ agrees well
with DIO, the procedure we have presented allows to place a bound on
$B(\mu\to e J)$. If it is found that the electron distribution is not
well described by DIO, it could be checked if the addition of a
component following Eq.~(\ref{eq:fitepAl}) improves the agreement
which could be consistent with the presence of events coming
from MEIO. Note that (when we neglect nuclear-recoil effects) the electron
spectrum for MEIO goes as $\sim(E_{\mu e}-E_e)^3$ near the endpoint,
while the usual muon DIO goes as $\sim(E_{\mu e}-E_e)^5$. MEIO is,
therefore, less suppressed than DIO in this region, and the endpoint is a favorable region to search for the $\mu\to e J$
process. To compare the shapes of the two processes near the endpoint, we also show in the second plot in
Fig.~\ref{fig:specAl} (Fig.~\ref{fig:specAu}) the electron spectrum
for DIO in Al (Au) as the
dashed line \cite{muonDIO}, multiplied by a constant $C=415$ ($C=333$) to make it coincide with the
MEIO value at $E_e=100$ MeV ($E_e=90$ MeV).

One may also ask which target materials give better sensitivity to
$B(\mu\to e J)$. When we increase $Z$, the fraction $f_J$ increases
(which improves the bound one would obtain), but the capture width and
$N_R$ also increase (which worsens the bound one would obtain). The
net result is that, for a given sensitivity to the conversion search,
the sensitivity to $B(\mu\to e J)$ is roughly independent of $Z$.

\section{Conclusions}\label{sec:concl}
We have computed the electron spectrum for the $\mu\to e J$ decay when
the muon is orbiting a nucleus. Using those results, we described
a procedure to probe the muon-electron-majoron coupling using $\mu \to e$
conversion experiments. Using results of conversion experiments to
improve the  limit on $B(\mu\to e J)$ does not require any
dedicated experimental search. $B(\mu\to e J)$ can also be
probed in
$\mu\to e\gamma$ searches \cite{Hirsch:2009ee}, although some
relaxation in the cuts of those experiments is required to improve the
present limits. The future
conversion experiments, Mu2e and COMET, may have the capability to produce bounds on
$B(\mu\to e J)$ that are competitive with the present ones, and possibly
improve them. The results
presented in this paper strengthen the physics case for the upcoming conversion experiments.

\begin{acknowledgments} This research was supported by the  Science and
Engineering Research Canada (NSERC). 
AC thanks the Institute for Nuclear Theory at
the University of Washington for its hospitality and the US Department of
Energy for partial support during the completion of this work. 

\end{acknowledgments}

\end{document}